\documentclass[aps,prb,twocolumn,superscriptaddress,showpacs]{revtex4-1}
\usepackage{amsmath,amssymb,amsfonts,amsthm}
\usepackage{graphicx}
\usepackage{bm}
\usepackage{bbm}
\usepackage{color}
\usepackage{dcolumn}   
\usepackage{epstopdf}
\usepackage[caption=false]{subfig}
\usepackage{color}
\usepackage[colorlinks=true,linkcolor=blue,urlcolor=blue,citecolor=blue]{hyperref}

\begin{document}

 \newcommand{\breite}{1.0} 

\newtheorem{prop}{Proposition}
\newtheorem{cor}{Corollary} 

\newcommand{\be}{\begin{equation}}
\newcommand{\ee}{\end{equation}}

\newcommand{\bea}{\begin{eqnarray}}
\newcommand{\eea}{\end{eqnarray}}
\newcommand{\lt}{<}
\newcommand{\gt}{>}

\newcommand{\Reals}{\mathbb{R}}     
\newcommand{\Com}{\mathbb{C}}       
\newcommand{\Nat}{\mathbb{N}}       

\newcommand{\id}{\mathbboldsymbol{1}}    

\newcommand{\Real}{\mathop{\mathrm{Re}}}
\newcommand{\Imag}{\mathop{\mathrm{Im}}}

\def\O{\mbox{$\mathcal{O}$}}   
\def\F{\mathcal{F}}			
\def\sgn{\text{sgn}}

\newcommand{\deo}{\ensuremath{\Delta_0}}
\newcommand{\dea}{\ensuremath{\Delta}}
\newcommand{\ak}{\ensuremath{a_k}}
\newcommand{\ad}{\ensuremath{a^{\dagger}_{-k}}}
\newcommand{\sx}{\ensuremath{\sigma_x}}
\newcommand{\sz}{\ensuremath{\sigma_z}}
\newcommand{\spl}{\ensuremath{\sigma_{+}}}
\newcommand{\smi}{\ensuremath{\sigma_{-}}}
\newcommand{\alk}{\ensuremath{\alpha_{k}}}
\newcommand{\bk}{\ensuremath{\beta_{k}}}
\newcommand{\ok}{\ensuremath{\omega_{k}}}
\newcommand{\vd}{\ensuremath{V^{\dagger}_1}}
\newcommand{\vi}{\ensuremath{V_1}}
\newcommand{\vo}{\ensuremath{V_o}}
\newcommand{\zc}{\ensuremath{\frac{E_z}{E}}}
\newcommand{\xc}{\ensuremath{\frac{\Delta}{E}}}
\newcommand{\xd}{\ensuremath{X^{\dagger}}}
\newcommand{\aok}{\ensuremath{\frac{\alk}{\ok}}}
\newcommand{\tpw}{\ensuremath{e^{i \ok s }}}
\newcommand{\tpe}{\ensuremath{e^{2iE s }}}
\newcommand{\tmw}{\ensuremath{e^{-i \ok s }}}
\newcommand{\tme}{\ensuremath{e^{-2iE s }}}
\newcommand{\epls}{\ensuremath{e^{F(s)}}}
\newcommand{\emis}{\ensuremath{e^{-F(s)}}}
\newcommand{\epl}{\ensuremath{e^{F(0)}}}
\newcommand{\emi}{\ensuremath{e^{F(0)}}}

\newcommand{\lr}[1]{\left( #1 \right)}
\newcommand{\lrs}[1]{\left( #1 \right)^2}
\newcommand{\lrb}[1]{\left< #1\right>}
\newcommand{\nbt}{\ensuremath{\lr{ \lr{n_k + 1} \tmw + n_k \tpw  }}}

\newcommand{\om}{\ensuremath{\omega}}
\newcommand{\dw}{\ensuremath{\Delta_0}}
\newcommand{\wbp}{\ensuremath{\omega_0}}
\newcommand{\dv}{\ensuremath{\Delta_0}}
\newcommand{\vbp}{\ensuremath{\nu_0}}
\newcommand{\vplus}{\ensuremath{\nu_{+}}}
\newcommand{\vminus}{\ensuremath{\nu_{-}}}
\newcommand{\wplus}{\ensuremath{\omega_{+}}}
\newcommand{\wminus}{\ensuremath{\omega_{-}}}
\newcommand{\uv}[1]{\ensuremath{\mathbf{\hat{#1}}}} 
\newcommand{\abs}[1]{\left| #1 \right|} 
\newcommand{\avg}[1]{\left< #1 \right>} 
\let\underdot=\d 
\renewcommand{\d}[2]{\frac{d #1}{d #2}} 
\newcommand{\dd}[2]{\frac{d^2 #1}{d #2^2}} 
\newcommand{\pd}[2]{\frac{\partial #1}{\partial #2}} 
\newcommand{\pdd}[2]{\frac{\partial^2 #1}{\partial #2^2}} 
\newcommand{\pdc}[3]{\left( \frac{\partial #1}{\partial #2}
 \right)_{#3}} 
\newcommand{\ket}[1]{\left| #1 \right>} 
\newcommand{\bra}[1]{\left< #1 \right|} 
\newcommand{\braket}[2]{\left< #1 \vphantom{#2} \right|
 \left. #2 \vphantom{#1} \right>} 
\newcommand{\matrixel}[3]{\left< #1 \vphantom{#2#3} \right|
 #2 \left| #3 \vphantom{#1#2} \right>} 
\newcommand{\grad}[1]{{\nabla} {#1}} 
\let\divsymb=\div 
\renewcommand{\div}[1]{{\nabla} \cdot \boldsymbol{#1}} 
\newcommand{\curl}[1]{{\nabla} \times \boldsymbol{#1}} 
\newcommand{\laplace}[1]{\nabla^2 \boldsymbol{#1}}
\newcommand{\vs}[1]{\boldsymbol{#1}}
\let\baraccent=\= 

\title{Quantum heat waves in a one-dimensional condensate}

\author{Kartiek Agarwal}
\email{kagarwal@princeton.edu}
\affiliation{Department of Electrical Engineering, Princeton University, Princeton, New Jersey 08540, USA}
\affiliation{Physics Department, Harvard University, Cambridge, Massachusetts 02138, USA}
\author{Emanuele G. Dalla Torre}
\affiliation{Department of Physics, Bar Ilan University, Ramat Gan 5290002, Israel}
\author{J\"org Schmiedmayer}
\affiliation{Vienna Center for Quantum Science and Technology, Atominstitut, TU Wien, Stadionallee 2, 1020 Wien, Austria}
\author{Eugene Demler}
\affiliation{Physics Department, Harvard University, Cambridge, Massachusetts 02138, USA}

\date{\today}
\begin{abstract}
We study the dynamics of phase relaxation between a pair of one-dimensional condensates created by a bi-directional, supersonic `unzipping' of a finite single condensate. We find that the system fractures into different \emph{extensive} chunks of space-time, within which correlations appear thermal but correspond to different effective temperatures. Coherences between different eigen-modes are crucial for understanding the development of such thermal correlations; at no point in time can our system be described by a generalized Gibbs' ensemble despite nearly always appearing locally thermal. We rationalize a picture of propagating fronts of hot and cold sound waves, populated at effective, relativistically red- and blue-shifted temperatures to intuitively explain our findings. The disparity between these hot and cold temperatures vanishes for the case of instantaneous splitting but diverges in the limit where the splitting velocity approaches the speed of sound; in this limit, a sonic boom occurs wherein the system is excited only along an infinitely narrow, and infinitely hot beam. We expect our findings to apply generally to the study of superluminal perturbations in systems with emergent Lorentz symmetry.
\end{abstract}
 
\maketitle

\section{Introduction} 

Coherent out-of-equilibrium dynamics of quantum many-body systems can often reveal a host of new phenomena that has no analog in equilibrium matter, and that demands an inquiry in itself. Bloch oscillations without a lattice~\cite{schecter2012dynamics}, superradiance~\cite{baumann2010dicke}, topological defect generation in quenches across phase transitions~\cite{sadler2006spontaneous}, topological phases induced by driving~\cite{kitagawa2011transport}, and non-vanishing infinite-time correlations in the many-body-localized phase~\cite{mblpalhuse,schreiber2015observation} are just some examples of such new phenomena. While such a regime has been challenging to achieve and reliably probe in traditional condensed matter systems (see however, Refs.~\cite{fausti2011light,wang2013observation,levonian2016probing,houck2012chip} for notable examples), which re-equilibrate on extremely short timescales due to strong coupling to the environment, gases of ultra-cold atoms can be operated as highly isolated~\cite{Greiner,Sidorov}, artificial quantum matter~\cite{Cheneau, Langen,Sadler,Chen,Ronzheimer,Trotzky} which can be assumed to evolve under its own dynamics over extremely long time-scales. This has reinvigorated interest in this line of inquiry, and has influenced many of the developments mentioned above. 

One-dimensional systems have, in particular, garnered attention due to their pronounced non-equilibrium behavior~\cite{Kinoshita06}, which arises due to the limited phase space for scattering and equilibration in these systems. In addition, most gapless one-dimensional systems, such as gases of interacting bosons or fermions, and many spin systems, exhibit an emergent Lorentz symmetry at low energies; this is efficiently captured by a description of the system as a collection of free bosons, and is known as the Luttinger Liquid theory~\cite{Haldane} (LLT). Besides observing the characteristic power-law decay of correlations in a one-dimensional Bose gases as predicted by LLT, non-equilibrium measurements have directly probed the linear dispersion of constituent excitations by observing the light-cone like spread of correlations~\cite{Cheneau,Langen}. 

An important, persistent line of inquiry has been to understand the re-equilibration process in isolated one dimensional systems~\cite{Meisner,Rigol2,Rigol3,Shashi,Cazalilla,Dalla,Trotzky}. While the integrable nature of the LLT rules out true equilibration, it has been predicted~\cite{Rigol} that these systems may enter a `pre-thermal' state described by a Generalized Gibbs' Ensemble (GGE) that correctly estimates the value of constants of motion of the non-equilibrated system. This is based on the idea that correlations associated with non-conserved operators, which in general are time-dependent, rapidly dephase and do not contribute significantly in most long-time measurements. When the constants of motion are limited to the population of various modes describable by a single temperature, the GGE is the usual Gibbs' Ensemble. A successful demonstration (among others~\cite{Kinoshita06,Jorgcross}) of these ideas was an experiment that observed the relaxation of the phase difference between two halves of a uniformly-split one-dimensional condensate~\cite{Gring}---the system was predicted to enter into a prethermal state~\cite{Bistritzer,Takuya} described by an effective temperature $T_0 = g \rho/2$, where $2 g \rho$ is the interaction energy density of the initial condensate, of density $2 \rho$---and experiments were able to verify these predictions. 

The GGE has proven to be a very useful tool in determining the long-time properties of many out-of-equilibrium systems~\cite{Rigol,CalabreseQuench,CalabreseTransverse2,IlievskispinlessGibbs,EsslerGibbsQFT,Gring,Jorgcross}. Given its successes, it is interesting to find and explore instances of quantum dynamics that go beyond the GGE paradigm, and where `off-diagonal' or time-dependent quantum coherences cannot be neglected, even at late times. In this paper, we provide precisely such an example, which allows a simple physical interpretation and which can be investigated experimentally. Perhaps more intriguingly, even though a successful description of our system requires taking into account these time-dependent quantum coherences, the system yet appears to be stationary and thermal on extensively (proportional to the system size) large regions of the space-time; these regions are correctly described by different effective temperatures, none of which agree with the temperature describing the occupation of the constants of motion.  

\begin{center}
\begin{figure}
\includegraphics[width = 2.8in]{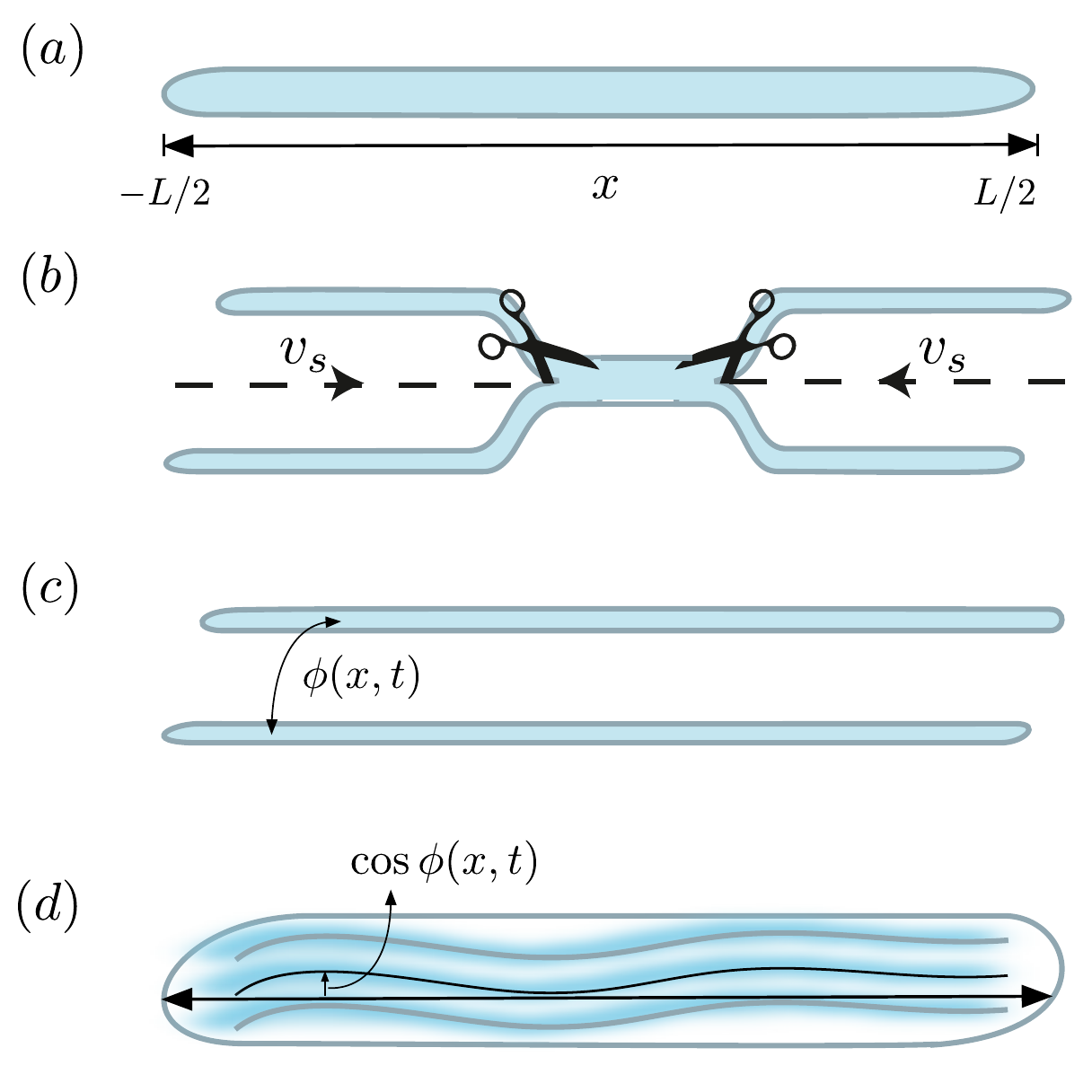}
\caption{Realization of the theoretical model using ultra-cold atoms. (a) A finite one-dimensional condensate, of length $L$, is prepared (at low temperatures) and (b) split into two halves along a supersonic `knife-edge' that travels at velocity $v_s$; (c) the phase difference between the halves $\phi(x,t)$ evolves as a Luttinger Liquid, and can be measured by interfering the two halves, as shown in (d).}
\label{fig:protocolfig}
\end{figure}
\end{center}

In particular, we consider a generalization of the condensate splitting experiments~\cite{Gring,Jorgcross} and propose studying phase relaxation dynamics after a splitting protocol in which the splitting occurs on two supersonic `knife-edges' that travel from either ends of the condensate towards the center; see Fig.~\ref{fig:protocolfig}. Such a perturbation differs from uniform, sudden quenches~\cite{CalabreseQuench,EsslerSpinChainQuenchReview}, smooth (potential) ramp protocols~\cite{Polkovnikov,BernierMottQuench,lamporesi2013spontaneous,KibbleBerkeley}, and semi-infinite quenches~\cite{InhomogeneousQPT} that have been previously employed to study non-equilibrium behavior in one-dimensional systems, and as we discuss, provides new insights into their dynamics, particularly in relation to their emergent Lorentz symmetry. The experimental setup for realizing such supersonic zippers has been discussed in Ref.~\cite{AgarwalChiral}. We also note that our analysis assumes that the condensate is uniform; such homogeneous systems have been realized using ultra-cold atoms in flat trapping potentials~\cite{HadzibabicUniformBose,mukherjee2015fermi}. 

\begin{figure}
\begin{center}
\includegraphics[width = 3 in]{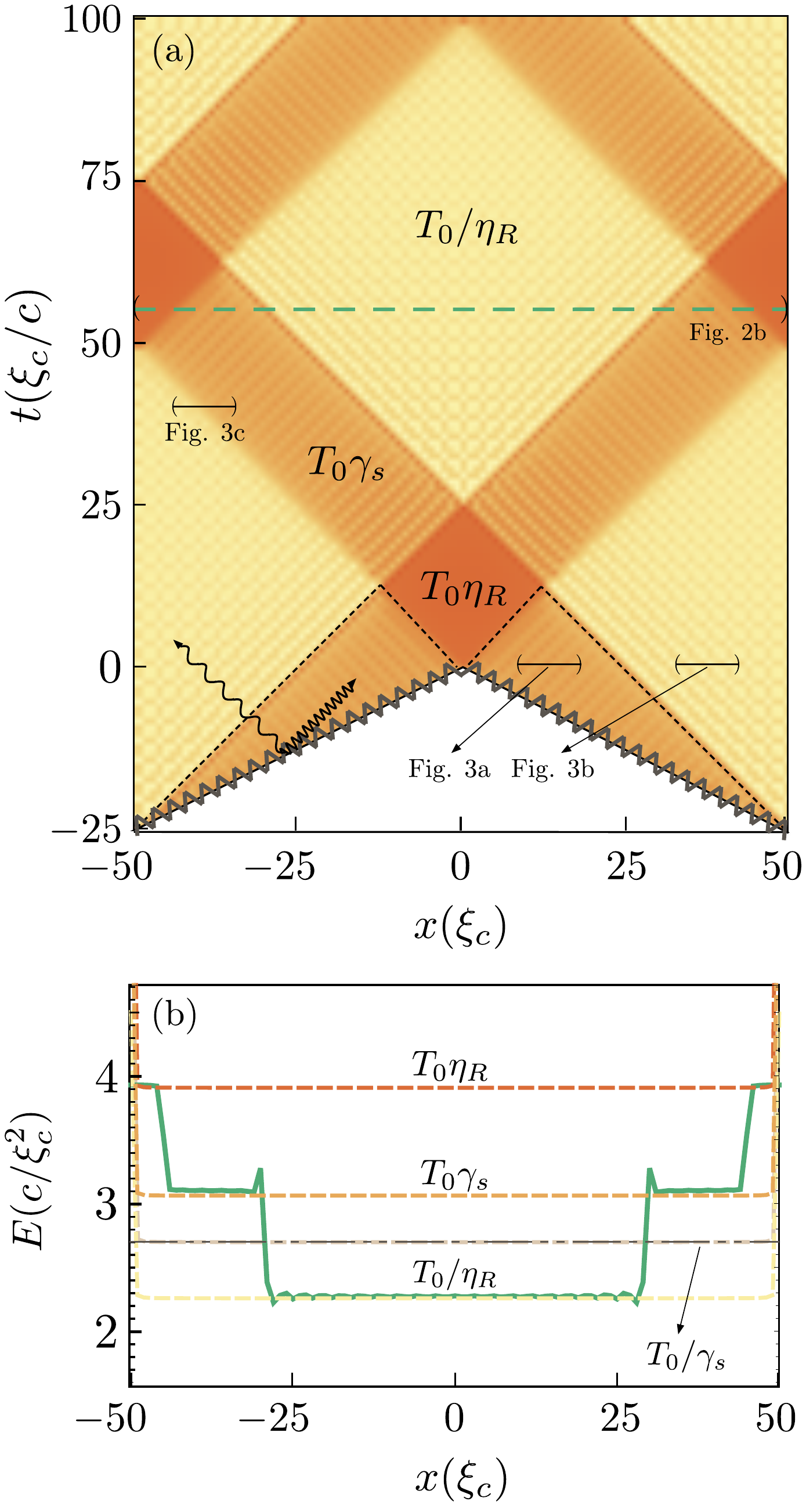} \\
\caption{(a) The energy density (large magnitude is shown in darker color) is plotted as a function of time on the y-axis (in units of $\xi_c/c$), and position on the x-axis (in units of $\xi_c$) for system-size $L = 100 \xi_c$, splitting velocity $v_s = 2c$, Luttinger parameter $K = \rho \xi_c /2 = 10$, and healing length $\xi_c$. Regions of different energy densities are created by waves emitted from the splitter at relativistically red- and blue- Doppler-shifted temperatures (curly lines emanating from splitter) are seen; these are described by a temperature $x T_0$, where $x$ is noted on the plot. The regions at temperature $T_0 \gamma_s$ immediately above the splitting front correspond to the regions described by previous work~\cite{AgarwalChiral}. We also indicate the regions considered in Figs.~\ref{fig:localtemp} (a), (b) and (c) to used the decay of phase correlations. (b) The energy density is plotted as a function of position at time $t = 60 \xi_c/c$ [green line, as also indicated in (a)]. Thick dashed lines correspond to energy density as expected from a system at temperature $xT_0$. The dash-dot line corresponds to the average energy density which is seen to coincide well with the energy density of the state at temperature $T_0/\gamma_s$.}
\label{fig:enerprop}
\end{center}
\end{figure}
\begin{figure}
\begin{center}
\includegraphics[width = 2.8 in]{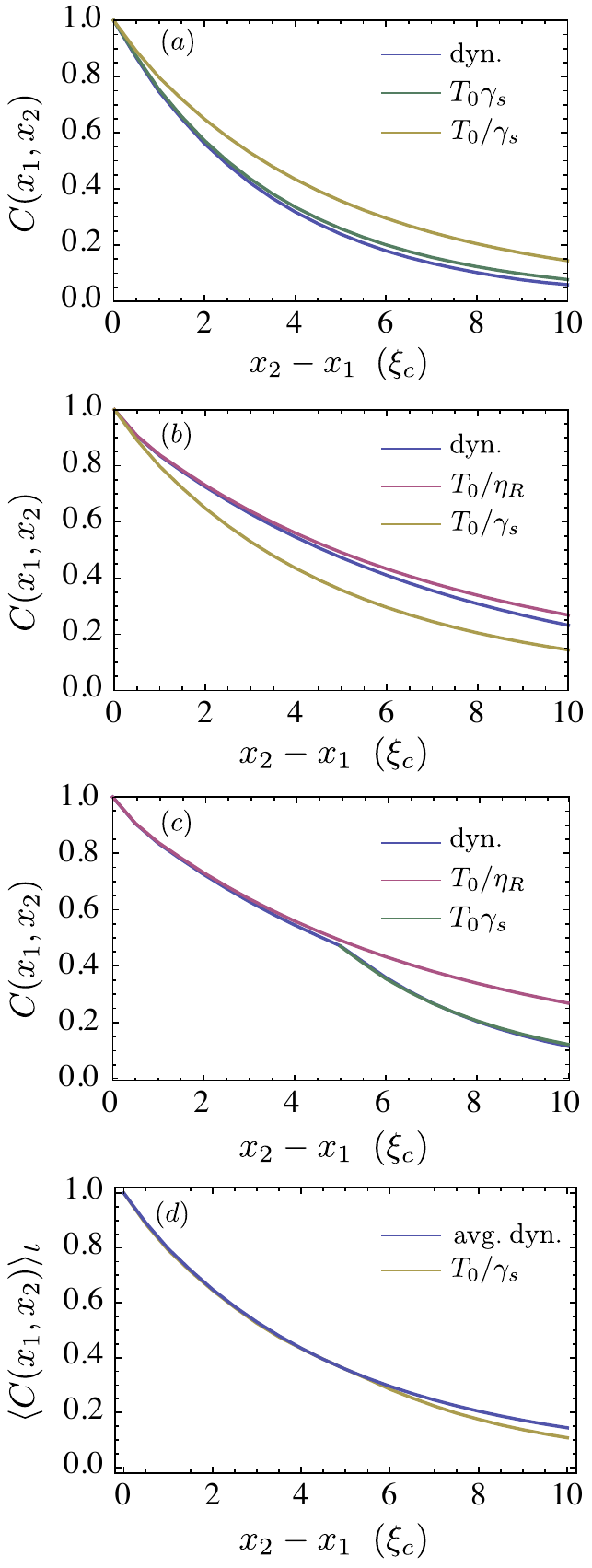} \\
\caption{Decay of equal-time correlations $C(x_1,x_2) = \text{exp} \left[-i (\phi(x_1) - \phi(x_2) )\right]$ in various space-time intervals; (a) $t = 0$, $x_1 = 5 \xi_c$; (b) $t = 0$, $x_1 = 35 \xi_c$; (c) $t = 37.5 \xi_c/c$, $x_1 = -42.5 \xi_c$; and (d) $t \in [0, L/c]$, $x_1 = -5 \xi_c$. $C(x_1,x_2,t)$ is log-averaged (see main text) over a complete evolution cycle, $t \in [0,L/c]$ and plotted in (d); it is seen to agree with the decay corresponding to a system at temperature $T_f = T_0/\gamma_s$, describing the occupation of LL bosons. In (a), (b) and (c), equal-time correlations calculated from a dynamical solution of the problem (blue) are seen to decay in space exponentially, and in agreement with the decay predicted by the local effective temperature of the region (red or green), as shown in Fig.~\ref{fig:enerprop};  these local temperatures do not agree with $T_f$. The space-time intervals considered in (a), (b) and (c) are also shown in Fig.~\ref{fig:enerprop}}
\label{fig:localtemp}
\end{center}
\end{figure}

We follow Refs.~\cite{Bistritzer,Takuya,AgarwalChiral}, and describe the phase-difference between the two `halves' of the condensate to be zero before the splitting and to evolve as a Luttinger Liquid (LL) thereafter. We utilize the Lorentz symmetry of the problem to present an analytical solution. Our analysis reveals that the system enters a state where the occupation of LL bosons is well approximated by a thermal distribution with temperature $T_f = T_0 / \gamma_s$, where $\gamma_s = 1/\sqrt{1 - c^2/v_s^2} >1$ is a Lorentz dilation factor associated with the inverse of the supersonic splitting velocity $v_s > c$. However, unlike most instances where the GGE applies, the occupation of these bosons is not sufficient to describe correlations of the system; it is important to consider the effect of off-diagonal correlations associated with LL bosons of different momenta, which in particular, are time-dependent. Nevertheless, we find that the dynamics of the system can be partitioned into various extensive space-time regions wherein correlations appear to be stabilized (in time), and are well described according to a `local' temperature---see Figs.~\ref{fig:enerprop} and~\ref{fig:localtemp}---and none of these local temperatures coincide with $T_f$.  

These space-time regions have a simple physical interpretation. The splitting protocol may be thought as one that generates two sets of bosonic excitations: one set propagates from the splitter along its direction of motion, and another set travels against it. In the Lorentz-boosted frame where the splitting is instantaneous, excitations can be expected to be populated at a temperature $T_0$ as per previous analyses~\cite{Bistritzer,Takuya}. In the laboratory frame, these excitations are Doppler shifted up or down by the relativistic Doppler factor $\eta_R = \sqrt{1+c/v_s}/\sqrt{1-c/v_s}$ depending on which direction they travel in (along or against the splitting trajectory). This creates wavefronts at temperatures $T_0/\eta_R$ (cold), and $T_0 \eta_R$ (hot), which can combine with one another to create regions at `local prethermal temperatures' given by $T_0/\eta_R$, $T_0 \eta_R$ or $T_0 \gamma_s = (T_0/\eta_R + T_0 \eta_R)/2$ depending on whether the region is populated only by cold, hot, or an equal admixture of cold and hot waves, respectively. [The space-time region above the splitter (enclosed by dashed lines in Fig.~\ref{fig:enerprop}), is described by a temperature $T_0 \gamma_s$, as predicted in a previous analysis which disregarded the finite size of the system, and assumed uni-directional splitting~\cite{AgarwalChiral}]. Thus, our system provides an interesting example where the GGE fails, and where correlations nevertheless become stationary (for extensively long periods of time), and appear thermal. 

We additionally find that correlation functions display `cross' like patterns (see Figs.~\ref{fig:cross} and~\ref{fig:cross2}), which indicate strong correlations of the relative phase at mirror-symmetric points $x$ and $-x$ (assuming the center of the condensate is at $x = 0$), and to some extent, mimic those already observed in experiments~\cite{Jorgcross}. In our analysis, these anomalous correlations arise due to a long-time quantum coherence phenomenon that first occurs at time $t_c (v_s) = L (1-c/v_s)/ (2 c)$ and recurs periodically thereafter (with time-period $L/c$). At $t_c$, the non-zero (anomalous) pair-correlation functions $\sim \avg{a^\dagger_{k_1} a^\dagger_{k_2}} (t) \sim e^{i (\abs{k_1} + \abs{k_2}) t} $(for some momenta $k_1$ and $k_2$), which normally dephase rapidly, are re-focussed. In particular, for the case of instantaneous splitting, these correlations (for $k_1 = k_2 = k$) have the same magnitude as the usual time-independent correlations proportional to the occupation number $\sim \avg{a^\dagger_{k} a_{k}}$ of excitations. Moreover, the phase $\cos (2 k t)$ inherited by these anomalous correlations at time $t = t_c$, is the same, $+1$, for \emph{all} symmetric modes (even functions in $x$), and $-1$ for \emph{all} anti-symmetric modes (odd functions of $x$). This leads to an ephemeral state where anti-symmetric modes appear to be in their ground state while symmetric modes appear to be populated at twice the temperature $T_f$. This difference of populations creates strong correlations of the phase at points $x$ and $-x$. We also discuss how these correlations spread at certain supersonic velocities $c'_\pm = 2c / (1 \pm u_s)$, providing both analytic and numerical confirmation of these ideas.

(We note that these anomalous correlations are generally not accessible experimentally due to rapid dephasing; thus, probing such revivals may allow for a more direct measurement of these correlations.)

Finally, we note that our analysis provides a general way to understand the non-equilibrium dynamics of finite Lorentz-invariant systems after an application of superluminal perturbations. Our results should apply, in general, to systems with quasiparticles governed by a linear dispersion, at times shorter than the dephasing time set by interactions between the quasiparticles. For inhomogeneous systems, such as those in a harmonic trap, our analysis is likely valid at time-scales smaller than the inverse trapping frequency~\cite{geiger2014local}, beyond which the heat waves should dephase due to the inhomogeneous sound velocity in the system. 

\section{Model and solution}

We follow Ref.~\cite{Takuya} and describe the condensate as a LL. The phase-difference $\phi$ between the `halves' of this single condensate can then also be described as a LL but with, additionally, an effective coupling, or mass that is set to zero along the splitting front. This coupling, with amplitude $J = g \rho^2$, suppresses the phase difference prior to the splitting, and is set in a way that guarantees the expected correlations at the time of the splitting~\cite{Bistritzer}, which is, $\avg{n(x), n(x')} = \rho/2 \delta(x-x')$ where $n(x)$ is the local density conjugate to the phase $\phi(x)$, $2 \rho$ is the density of the undivided condensate, and $\delta(x)$ is the Dirac delta-function defined over the length-scale of the healing length $\xi_c = \pi/\sqrt{m_b \rho g}$; $m_b$ is the mass of the bosons and $g$ is the effective point-scattering amplitude. These correlations can be justified by estimating the boson number fluctuations between the halves of the condensate~\cite{Bistritzer}. The Hamiltonian describing a splitter or `mass front' traveling from from either side inwards to the center $x = 0$ at velocity $v_s$ reads

\begin{align}
H(t) &= \int_{-L/2}^{L/2} dx \; g n^2 + \frac{\rho}{4m_b} (\partial_x \phi)^2 + J (x,t) \phi^2, \nonumber \\
J(x,t) &= g \rho^2 \left(1-\Theta(x +v_s t) \right) \left( \Theta(x-v_s t)  \right). 
\end{align}

Here we have defined time such that the splitting protocol begins at time $t = - L/(2v_s)$ and is complete at $t = 0$. Before we analyze the dynamics in more detail, we note that, if $J$ were fixed in time at $J = g \rho^2$, the dynamics of the system would be massive, $(\partial^2_t - c^2 \partial^2_x + m^2) \phi = 0$, with a (energies above which the Luttinger description fails) mass $m = 2 g \rho$ which is the chemical potential of the undivided condensate. This value of this mass is also the UV cut-off associated with the Luttinger theory; thus, it quenches relative phase fluctuations over the healing length, as is natural for a single condensate. It also implies that one can assume the `system' describing the relative phase and density fluctuations is at zero temperature even though the condensate itself may be at some finite temperature. (This has also been experimentally observed~\cite{Gring}.) In what follows, we set $c = \xi_c = 1$.

\subsection{Dynamics and Boundary Conditions} 

To avoid dealing with two mass fronts and make our problem more tractable, we separately analyze the Hamiltonian for symmetric $(+)$ and anti-symmetric $(-)$ combinations of the field $\phi$ defined for $x>0$--$\phi_{\pm}(x) = (\phi(x) \pm \phi(-x))/2$. The original field $\phi$ satisfies $\phi(x>0) = \phi_+(x) + \phi_-(x)$ and $\phi(x<0) = \phi_+(|x|) - \phi_- (|x|)$. Similarly, we define conjugate fields $n_{\pm} = n(x) \pm n(-x)$  with $n = (n_+ \pm n_-)/2$ depending on the sign of $x$. 

The Hamiltonians $H_{\pm}$ for fields $\phi_{\pm}$, are decoupled and read 

\begin{align}
H_{\pm} &= \int^{L/2}_0 dx \; \frac{g}{2} n^2_{\pm} + \frac{\rho}{2m_b} (\partial_x \phi_{\pm})^2 + J_\pm (x,t) \phi_{\pm}^2;   \nonumber \\
J_\pm (x,t) &= 2 g \rho^2 \left(1-\Theta(x +v_s t) \right). 
\end{align}

and the fields obey the boundary conditions $\phi_- (0) = 0$, $\partial_x \phi_+ (0) = 0$ at x = 0. The boundary conditions at $x = L/2$ require careful consideration. Generally, we expect for a theory of weakly interacting bosons, that the boson current vanishes at the edge of the condensate; in this case, $\partial_x \phi_\pm (L/2) = 0$. If, on the other hand, the system is strongly interacting, it may be more meaningful to pin the density of bosons at the edge as zero, in which case, $n \sim \partial_t \phi_\pm (L/2) = 0$. We note that these boundary conditions are not crucial for thermodynamic results but play a role in deciding the form of late-time quantum coherences. We provide a description of both these cases in what follows. 

\begin{center}
\begin{figure}
\includegraphics[width = 3.35 in]{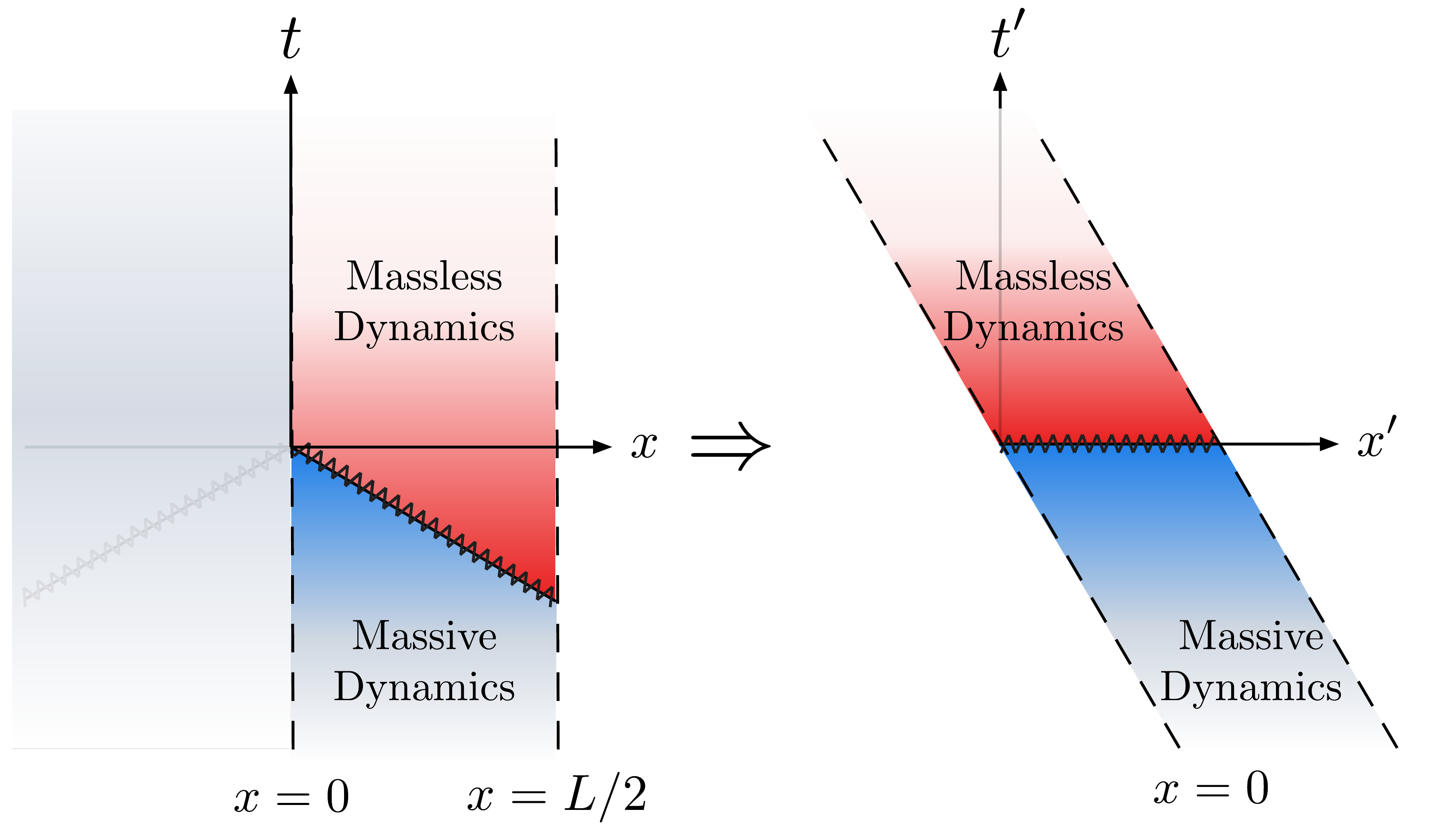}
\caption{Comparison of dynamics and boundary conditions in the laboratory frame and Lorentz boosted frame }
\label{fig:spacetime}
\end{figure}
\end{center}

Following Ref.~\cite{AgarwalChiral}, we work in a Lorentz-boosted frame with coordinates $(x',t')$ related to $(x,t)$ by the relations $x' = \gamma_s (x-u_s t)$, $t' = \gamma_s (t - u_s x / c^2)$, and $\gamma_s = \frac{1}{\sqrt{1-(u_s/c)^2}} = \frac{1}{\sqrt{1-(c/v_s)^2}}$. Note that $u_s$, the velocity of the Lorentz boost, is given by $u_s = c^2/v_s < c$. Under this transformation, the quench becomes homogeneous: $1-\Theta(x +v_s t)= \Theta (- t')$. Note that, in the boosted frame, the boundary conditions now need to be enforced on moving trajectories, $x' - u_s t'  = 0 \equiv x = 0$ and $x' - u_s t' = \frac{L}{2 \gamma_s}\equiv x = L/2$. In summary, 

\begin{align}
1-\Theta(x +v_s t) &\rightarrow \Theta (- t'), \nonumber \\
\phi_A (x = 0) = 0 &\rightarrow \phi_A (x' = u_s t') = 0, \nonumber \\
\partial_x \phi_S (x = 0) = 0 &\rightarrow  (\partial_x' + u_s \partial_t') \phi_S (x' = u_s t') = 0,
\label{eq:boundcond}
\end{align}

and similarly for the boundary conditions at $x = L/2$. The Lorentz transformation is illustrated in a space-time diagram in Fig. (\ref{fig:spacetime}). 

Note that the equations of motion for the modes for $t'<0$ in the Lorentz boosted frame are $\partial^2_{t'} \phi_\pm - c^2 \partial^2_{x'} \phi_\pm + m^2 \phi_\pm = 0$ where $m = 2 g \rho$ clearly acts as a `mass' of the constituent particles. For $t'>0$, this mass is set to 0. Thus, in the Lorentz-boosted frame, we have to solve the problem of a instantaneous quench, where the modes in our system go from being massive to massless, but have somewhat complicated boundary conditions. In the next section, we provide the complete set of eigenmodes for this problem.

\subsection{Eigenmodes in the Lorentz-boosted and laboratory frames}

In time-independent quantum mechanics, the orthogonality condition of various eigenstates reads $\int dx \; \phi_{k_1}^* (x) \phi_ {k_2} (x) = \delta(k_1 - k_2)$. However, this inner product does not encapsulate a fundamental feature of our problem--that of Lorentz symmetry. An inner product that is invariant under Lorentz transformations is the Klein-Gordon (KG) inner product~\cite{Birrell}, which is the Wronskian

\be
(u_{k_1}, u_{k_2} ) = -i \int_0^{L/2} dx \left( u_{k_1} \partial_t u^*_{k_2} - u^*_{k_2} \partial_t u_{k_1} \right) = g \delta (k_1 - k_2).
\label{eq:norm}
\ee

The KG inner product is invariant under Lorentz transformations in the sense that the integral over any space-like hyper-surface preserves the norm. It has additional symmetries associated with complex conjugation; $(a,b)^* = (b,a)$ and $(a,b)^* = - (a^*,b^*)$. Note that choice of the an overall factor in the norm is arbitrary; we set it to $g$. 

We now label the solutions of the equations of motion of $\phi_\pm(x',t')$ for $t' < 0$ by $v^{\pm}_k$, and for $t'>0$ by $u^\pm_k$, where $k$ will be an index associated with the momentum of the waves composing the solution. To derive these modes, we consider symmetric and anti-symmetric combinations of a pair of left- and right- moving waves; we set the left-moving wave to have momentum $k$ and energy $\omega_k = k$ (massless) or $\omega'_k = \sqrt{m^2 + k^2}$ (massive) and choose the momentum of the right-moving wave in a way that satisfies the boundary conditions at $x=0$. The symmetric $(+)$ and anti-symmetric $(-)$, forward-propagating (with a positive energy) modes in the boosted frame are given by

\begin{align} 
u^{\pm}_k &= A_k \left( e^{- i k (x' + t')} \pm e^{i \eta_R^2 k (x' - t')} \right) \; \; \; \text{for} \; \; k \ge 0, \nonumber \\
v^{\pm}_k &= B_k \left( e^{-i k x - i \omega'_k t } \pm e^{i f(k) x - i \omega'_{f(k)}} t \right) \; \; \; \text{for} \; \; k \ge - k_0, \nonumber \\
\text{where} \nonumber \\
A_k &= \sqrt{\frac{g}{2 L \eta_R k}}, \nonumber \\
B_k &= \sqrt{\frac{g}{2 L \gamma_s (k u_s + \omega'_k)}}, \nonumber \\
\eta_R^2 &= \frac{1+u_s}{1-u_s}, \nonumber \\
f(k) &= \frac{1+u^2_s}{1-u^2_s} k + \frac{2 u_s}{1-u^2_s} \omega'_k, \nonumber \\
k_0 &= m \frac{u_s}{\sqrt{1-u^2_s}}, \nonumber \\
\label{eq:basesboost}
\end{align}

Note that since we isolated symmetric and anti-symmetric modes, it is sufficient to consider only the modes with momentum $k \ge 0$ (for $t' \ge 0$) and $k \ge -k_0$ for $t'< 0$. 

(The choice $k \ge -k_0$ for the massive modes may appear confusing. Note that the momentum of the right-moving wave, $f(k)$, satisfies $f(-k_0) = k_0$ and that $f(k > -k_0) > k_0$. Thus, the choice $k \ge -k_0$ allows for right-moving waves of all momenta to be considered. Another way to justify this is by noting that for $k \ge -k_0$, eigenmodes in the laboratory frame exhibit all positive momenta.)

The normalization factors $A_k$, $B_k$ are found by setting the correlation $\avg{n(x) n(x')} = \frac{\rho}{2} \delta (x - x')$, and using the fact that the modes $v^\pm_k$ are in their ground state. To be precise, we note that $n(x) = \dot{\phi}/(2g)$, and that $\avg{n(x) n(x')} = \sum_{k \ge -k_0, \epsilon} \frac{1}{4g^2} v^\epsilon_{k} (x) v*^\epsilon_{k} (x')$; we approximate the integral $\int^\infty_0 \frac{dk}{2 \pi} \sqrt{m^2 + k^2} \frac{1}{g} \cos [k (x-x')] \approx \frac{\rho}{2} \delta (x-x')$. The approximation is valid at length scales $l \gg \xi_c$ where $\xi_c$ is the healing length = $\pi/\sqrt{m_b g \rho}$, where $m_b$ is the actual mass of the bosons forming the condensate. We also note that, fixing the normalization $A_k$, and $B_k$, determines the normalization factor $g$ in the KG inner product [in Eq.~(\ref{eq:norm})].

Finally, we apply the boundary condition at $x = L/2$; this finally determines the possible values of the momenta $k$ and is most easily done by examining the modes in the laboratory frame: 

\begin{align}
u^{-}_k &= i \sqrt{\frac{2g}{L \eta_R k}} \;   \sin{ \eta_R k x} e^{-i \eta_R k t} \; \; k =  \frac{ 2n \pi }{\eta_R L}, n \ge 0  \nonumber \\
u^{+}_k &=  \sqrt{\frac{2g}{L \eta_R k}} \; \cos{ \eta_R k x} e^{-i \eta_R k t} \; \; k = \frac{2 n \pi + \pi}{\eta_R L}, n > 0   \nonumber \\
v^{-}_k &=  i \sqrt{\frac{2g}{L \gamma_s (k u_s + \omega'_k)}} \; \sin{ \gamma_s (k + u_s \omega'_k)} e^{- i \gamma_s (k u_s + \omega'_k) t} \nonumber \\
& \gamma_s (k + u_s \omega'_k) L/2 = (2 n + 1 ) \pi/2 , n \ge 0 \nonumber \\
v^{+}_k &=  \sqrt{2 \frac{g}{L \gamma_s (k u_s + \omega'_k)}} \cos{ \gamma_s (k + u_s \omega'_k)} e^{- i \gamma (k u_s + \omega'_k) t} \nonumber \\
& \gamma (k + u_s \omega'_k) L/2 = n \pi, n > 0 \nonumber \\
v^{+}_{-k_0} &= \sqrt{\frac{g}{L m}}
\label{eq:baseslab}
\end{align}

where $n$ is an integer. In the above, we mention the quantized values of the momenta for the boundary condition $ \partial_t \phi(L/2) = 0$. If, alternatively, we use the boundary condition $\partial_x \phi(L/2) = 0$, the momentum values of symmetric and anti-symmetric modes is exchanged. While this may seem an unimportant detail, we note that, different boundary conditions change the nature of late-time quantum coherences that give rise to anomalous space-like correlations. We will return to this point in Sec.~\ref{sec:crosscorrsec}. 

We note that for $k = -k_0$, the combination $\gamma_s (k + u_s \omega'_k)$ appearing in harmonic modes $v^\pm_k$ is, in fact, zero. This again justifies our choice of a complete basis with $k \ge -k_0$ as mentioned previously. In particular, the symmetric zero mode $v^+_{-k_0}$ must be chosen with an extra factor of $1/\sqrt{2}$ compared to the simple analytic continuation of $v^+_{k>-k_0}$ in order to get correction normalization and commutation relations for the finite-size system. 

\subsection{Bogoluibov coeffecients}

For $t'<0$, the phase $\phi(x',t') =  \sum_{\epsilon = \pm}  \sum_{k \ge -k_0} \left( b^{\epsilon}_k v^{\epsilon}_k (x',t') + b^{\epsilon \dagger}_k v^{\epsilon*}_k (x',t') \right)$. We assume that initially the system is in the ground state of these modes, so that $\avg{b^\dagger_k b_k} = 0$. For $t' > 0$, it is useful to express the phase $\phi(x',t')$ in terms of the new eigenmodes $u^\pm_k$, that is, $\phi(x',t') = \sum_{\epsilon = \pm} \sum_k \left( a^{\epsilon}_k u^\epsilon_k (x',t') + a^{\epsilon \dagger}_k u^{\epsilon *}_k (x',t') \right)$. 

The relation between the original eigenmodes $v^\pm_k$ and the new eigenmodes $u^\pm_k$ can be expressed in terms of the Bogoluibov coefficients $\alpha^\epsilon_{k,k'}$ and $\beta^\epsilon_{k,k'}$ which are defined by the following relations

\begin{align}
\alpha^\epsilon_{k,k'} &= \frac{1}{g} ( u^\epsilon_k, v^\epsilon_{k'} ), \nonumber \\
\beta^\epsilon_{k,k'} &= -\frac{1}{g} ( u^\epsilon_k, v^{\epsilon*}_{k'} ), 
\end{align}

such that, 

\begin{align}
u^\pm_k &= \sum_{k'} \alpha^\pm_{k k'} v^{\pm}_{k'} + \beta^\pm_{k k'} v^{\pm*}_{k'}, \nonumber \\
v^\pm_{k'} &= \sum_{k} \alpha^{\pm*}_{k k'} u^{\pm}_{k} - \beta^\pm_{k k'} u^{\pm*}_{k}, \nonumber \\
a^{\pm}_k &= \sum_{k'} \alpha^{\pm*}_{k k'} b^\pm_{k'} - \beta^{\pm*}_{k k'} b^{\pm \dagger}_{k'}.
\end{align}

The explicit expressions for $\beta^\epsilon_{k k'}$  and $\alpha^\epsilon_{k k'}$ are

\begin{widetext}
\begin{align}
\beta^\pm_{k,k'} &= \frac{1}{g} A_k B_{k'} \frac{L}{2 \gamma_s} \Bigg\{ (\omega'_{k'} - k) e^{-i (k+k') \frac{L}{4 \gamma_s}} \text{sinc} \left[ (k+k') \frac{L}{4 \gamma_s} \right] \pm (\omega'_{f(k')} - k) e^{i (f(k')-k) \frac{L}{4 \gamma_s}} \text{sinc} \left[ (f(k')-k)\frac{L}{4 \gamma_s} \right]    \nonumber \\
&\pm (\omega'(k') - \eta_R^2 k) e^{i (\eta_R^2 k - k') \frac{L}{4 \gamma_s}} \text{sinc}\left[(\eta_R^2 k - k') \frac{L}{4 \gamma_s} \right] + (\omega'_{f(k')} - \eta_R^2 k) e^{i (f(k') + \eta_R^2 k) \frac{L}{4 \gamma_s}} \text{sinc} \left[ ( f(k') + \eta_R^2 k ) \frac{L}{4 \gamma_s} \right]  \Bigg\} \nonumber \\
\alpha^\pm_{k,k'} &= \frac{1}{g} A_k B_{k'} \frac{L}{2 \gamma_s} \Bigg\{ (\omega'_{k'} + k) e^{-i (k-k') \frac{L}{4 \gamma_s}} \text{sinc} \left[ (k'-k)\frac{L}{4 \gamma_s} \right]  \pm (\omega'_{f(k')} + k) e^{-i (f(k')+k) \frac{L}{4 \gamma_s}} \text{sinc} \left[ (f(k')+k)\frac{L}{4 \gamma_s} \right]   \nonumber \\
&\pm (\omega'(k') + \eta_R^2 k) e^{i (\eta_R^2 k + k') \frac{L}{4 \gamma_s}} \text{sinc} \left[(\eta_R^2 k + k') \frac{L}{4 \gamma_s} \right] + (\omega'_{f(k')} + \eta_R^2 k) e^{-i (f(k') - \eta_R^2 k) \frac{L}{4 \gamma_s}} \text{sinc} \left[( f(k') - \eta_R^2 k ) \frac{L}{4 \gamma_s} \right] \Bigg\}
\label{eq:alphbet}
\end{align}
\end{widetext}

A relation, $\sum_{k'} \alpha^{\pm*}_{k_1 k'} \alpha^\pm_{k_2 k'} - \beta^{\pm*}_{k_1 k'} \beta^\pm_{k_2 k'} = \delta_{k_1 k_2}$, exists due to the commutation relations: $[ a^\pm_k, a^{\pm \dagger}_{k'} ]  = \delta (k - k')$. These relations provide an important numerical check on the simulation of the results in the following sections.  

\subsection{Population of states}

The population of various $u^\epsilon_{k}$ modes, $N^\epsilon_k = \avg{a^{\epsilon \dagger}_k a^\epsilon_k}$ is given by 

\begin{align}
N^\epsilon_k = \sum_{k'} \abs{\beta^\epsilon_{k k'}}^2
\label{eq:pop}
\end{align}

An exact expression for $N^\epsilon_k$ can be given in the thermodynamic limit, $L \rightarrow \infty$. In this limit, $\frac{L}{2 \gamma_s} \text{sinc} (\frac{x L}{4 \gamma_s})$ is an approximation of $2 \pi \delta(x)$, where $\delta(x)$ is the Dirac delta-function. $\delta(0) = \frac{L}{4 \pi \gamma_s}$ is finite but tends towards infinity. We can now evaluate integrals such as $\int dx \; f(x) \delta^2(x) = f(0) \delta(0)$, and find for $L \rightarrow \infty$, 

\begin{align}
\beta^\epsilon_{k k'} &= \frac{2 \pi}{g} A_k B_{k'} \big\{ (\omega'_{k'} - k) [\delta (k' + k) + \delta(f(k') - k)] \nonumber \\
&+ \epsilon (\omega'_{\eta_R^2 k} - \eta_R^2 k) \delta (k' - \eta_R^2 k) \big\}, \nonumber \\
\alpha^\epsilon_{k k'} &= \frac{2 \pi}{g} A_k B_{k'} \big\{ (\omega'_{k'} + k) \delta (k' - k) \nonumber \\
&+ (\omega'_{\eta_R^2 k} + \eta_R^2 k) [\epsilon \delta (k' + \eta_R^2 k) + \delta(f(k') - \eta_R^2 k)]\big\}.
\label{eq:alphabetathermo}
\end{align}

The population of symmetric and anti-symmetric modes can be evaluated straightforwardly using Eqs.~(\ref{eq:pop}) and (\ref{eq:alphabetathermo}) 

\begin{align}
N^\epsilon_k &= \frac{1}{8 \gamma_s} \frac{1}{\eta_R k} \left[ \frac{(\omega'_k - k)^2}{\omega'_k} + \frac{(\omega'_{\eta_R^2 k} - \eta_R^2 k)^2}{\omega'_{\eta_R^2 k}} \right]
\label{eq:popu}
\end{align}

In finding this result, we note that only two of the delta functions in Eq.~\ref{eq:alphabetathermo} evaluated to non-zero values: these where the first and third delta functions for $k < k_0$ and the first and second delta functions for $k > k_0$. As it turns out, the result for $N^\epsilon_k$ is, nevertheless, a continuous function of $k$. We note that both anti-symmetric and symmetric modes have identical populations in the thermodynamic limit.

It is possible to approximately capture the occupation of these modes~\cite{mitra2011mode,dallatorre2013keldysh} by defining an effective temperature $T_f$ by noting that for bosons with an energy $\eta_R k$, the equilibrium distribution at temperature $T \gg \eta_R k$ is $N_k \approx T / ( \eta_R k)$. Using Eq.~(\ref{eq:popu}), we obtain an effective temperature $T_f \equiv \lim_{k \rightarrow 0} \eta_R k N^\epsilon_k = \frac{g \rho}{2 \gamma_s} = \frac{T_0}{\gamma_s}$.

\section{Failure of GGE and local prethermalization}

In this section, we discuss the failure of the Generalized Gibbs' ensemble in describing the correlations in our system. This fact is most clearly seen in the numerical simulations of the evolution of the energy density in time, as shown in Fig.~\ref{fig:enerprop} and the failure of the temperature $T_f$ in describing the decay of spatial correlations, as shown in Fig.~\ref{fig:localtemp}. In particular, we see in Fig.~\ref{fig:enerprop} that different space-time regions have clearly disparate energy densities; within these space-time regions, the energy is spread relatively uniformly. As described in the figure and as we will discuss below, we can provide an intuitive explanation of these individual regions in terms of the hot and cold wave fronts that emanate from the splitter. This picture also suggests that the local correlations within these space-time regions can be described by different effective temperatures, and this expectation is verified in Fig.~\ref{fig:localtemp} (a), (b) and (c). First, however, we explain how the temperature $T_f$, describing the occupation of the excitations, appears in our picture of these heat fronts, and how it can describe certain time-averaged correlations, as shown in Fig.~\ref{fig:localtemp} (d). Concomitantly, we explain why the GGE, $\rho_{\text{GGE}} \sim \text{exp} \left( - \sum_{k,\epsilon} a^{\pm \dagger}_k a^\pm_k / N^\epsilon_k  \right) \approx \text{exp} \left( - \sum_{k,\epsilon} E_k a^{\pm \dagger}_k a^\pm_k / T_f  \right)$, where $E_k = \eta_R k$ is the energy of the mode $a^\pm_k$, fails in describing the local correlations of the system. Finally, we explain how the effective temperatures of different space-time regions, as shown in Fig.~\ref{fig:enerprop}, are evaluated.

\subsection{Intuitive explanation of the effective temperature}

We now rationalize the result $T_{k \rightarrow 0} = \frac{g \rho}{2 \gamma_s}$. To this end, let us first note that, for the instantaneous splitting case ($\gamma_s = 1$, $v_s = \infty$), the system is known to prethermalize at a temperature $T_0 = g \rho/2$. This is also confirmed by our current analysis, since $\gamma_s = 1$ for $v_s = \infty$. 

We can expect this result for the sudden-quench to hold in the Lorentz-boosted frame where the quench is also sudden and uniform. Thus, in this frame, we must also have right- and left-moving waves with an average energy captured by the temperature $T_0$. Back in the laboratory frame, these waves are blue- or red-shifted by a factor of $\eta_R$, if they travel with or against, respectively, the splitting front. For instance, we expect the right-moving waves coming from the right-half splitter ($x > 0$) to be at a temperature $T_0 / \eta_R$ while the left-moving waves to have an effective temperature of $T_0 \eta_R$ (see Fig.~\ref{fig:enerprop}); the Doppler-shifting of the temperature follows from the Doppler-shift of the momenta and the fact that the energy of the waves is linear in the momentum. 

Next, we note that the width of the wave-front (at any fixed time $t$) of the hot waves is $L (1 - u_s)$ while that of the cold waves is $L (1 + u_s)$; the sum of these is of course, is $2 L$, with the factor of two corresponding to the existence of right- and left-moving waves. The average temperature of the system can then be computed to be $T_f = [T_0 \eta_R \times (1 - u_s) + T_0 / \eta_R \times (1 + u_s)]/2 = T_0 / \gamma_s$, which agrees with the finding that $T_{k \rightarrow 0} = \frac{g \rho}{2 \gamma_s}$. 

We now discuss how this effective temperature can describe certain time-averaged correlations but not fixed-time correlations.

\subsection{Time-averaged prethermalization}

 If we express $\phi(x,t) = \sum_{\epsilon = \pm} \sum_k \left( a^\epsilon_k u^\epsilon_k (x,t) +  a^{\epsilon \dagger}_k u^{\epsilon *}_k (x,t) \right)$, correlations of the form $\avg{\phi(x_1,t) \phi(x_2,t)}$ depend on averages of the sort $\avg{a^{\pm \dagger}_{k_1} a^\pm_{k_2}}$ which come with a time evolution $e^{-i \eta_R (k_1 - k_2) t}$, and $\avg{a^\pm_{k_1} a^\pm_{k_2}}$ which come with a time evolution $e^{\pm i \eta_R (k_1 + k_2) t}$. In particular, for equal time-correlations, we find the two-point correlator
 
\begin{align}
& \avg{[ \phi(x_1) - \phi(x_2) ]^2} (t) = \nonumber \\
& \sum_{} (2 \beta^\epsilon_{k_1 k} \beta^{\epsilon*}_{k_2 k} + 1) \left(u^\epsilon_{k_1} (x_1) - u^\epsilon_{k_1} (x_2) \right) \left( u^{\epsilon*}_{k_2} (x_1) - u^{\epsilon*}_{k_2} (x_2)  \right) \nonumber \\
&- \sum_{} \alpha^{\epsilon*}_{k_1 k} \beta_{k_2 k}^{\epsilon*} \left(u^\epsilon_{k_1} (x_1) - u^\epsilon_{k_1} (x_2) \right) \left( u^\epsilon_{k_2} (x_1) - u^\epsilon_{k_2} (x_2)  \right) \nonumber \\
&+ c.c.,
\end{align} 
 
where the sum is over the indices $k_1, k_2, k, \epsilon$. For large times $t \sim \mathcal{O} [L/c]$, we expect that the anomalous correlations, which come with a time evolution of the form $e^{\eta_R (k_1 + k_2) t}$ quickly dephase and do not contribute to the sum. 

Furthermore, we note that the truly time-independent part of the correlations comes from terms with $k_1 = k_2$, that is, with prefactors $| \beta^\epsilon_{k_1 k} |^2$. Thus, using the definition of the population of modes $N^\epsilon_k$ in Eq.~(\ref{eq:pop}), we see that the time-independent part of the correlations of the system is given by  

\begin{align}
\avg{[ \phi(x_1) - \phi(x_2) ]^2} & (t \sim \mathcal{O} [ L/c] ) & \nonumber \\
\approx & \sum_{\epsilon, k} (2 N^\epsilon_k + 1) \abs{u^\epsilon_k (x_1) - u^\epsilon_k (x_2) }^2. 
\end{align}

The above time-independent correlations can be reproduced by assuming that the system is in an excited state with a population given by $N^\epsilon_k$ as given in Eq.~(\ref{eq:popu}), or, equivalently, by assuming that the system is at the temperature $T_f$. Assuming all other correlators can be neglected, we may expect that the system should reach a prethermal state where equal-time correlations do not depend on the time. Moreover, since the theory is quadratic, all correlation functions can be decomposed into two-point correlations and thus, retain the thermal aspect of the correlations; one may also construct distribution functions $P_l (\alpha)$ of the phase contrast $\alpha = \abs{\int^l_0 dx e^{-i \phi(x,t)}}^2$ as carried out in previous works~\cite{Imambekov,Takuya}.

\begin{center}
\begin{figure*}
\includegraphics[width = \textwidth]{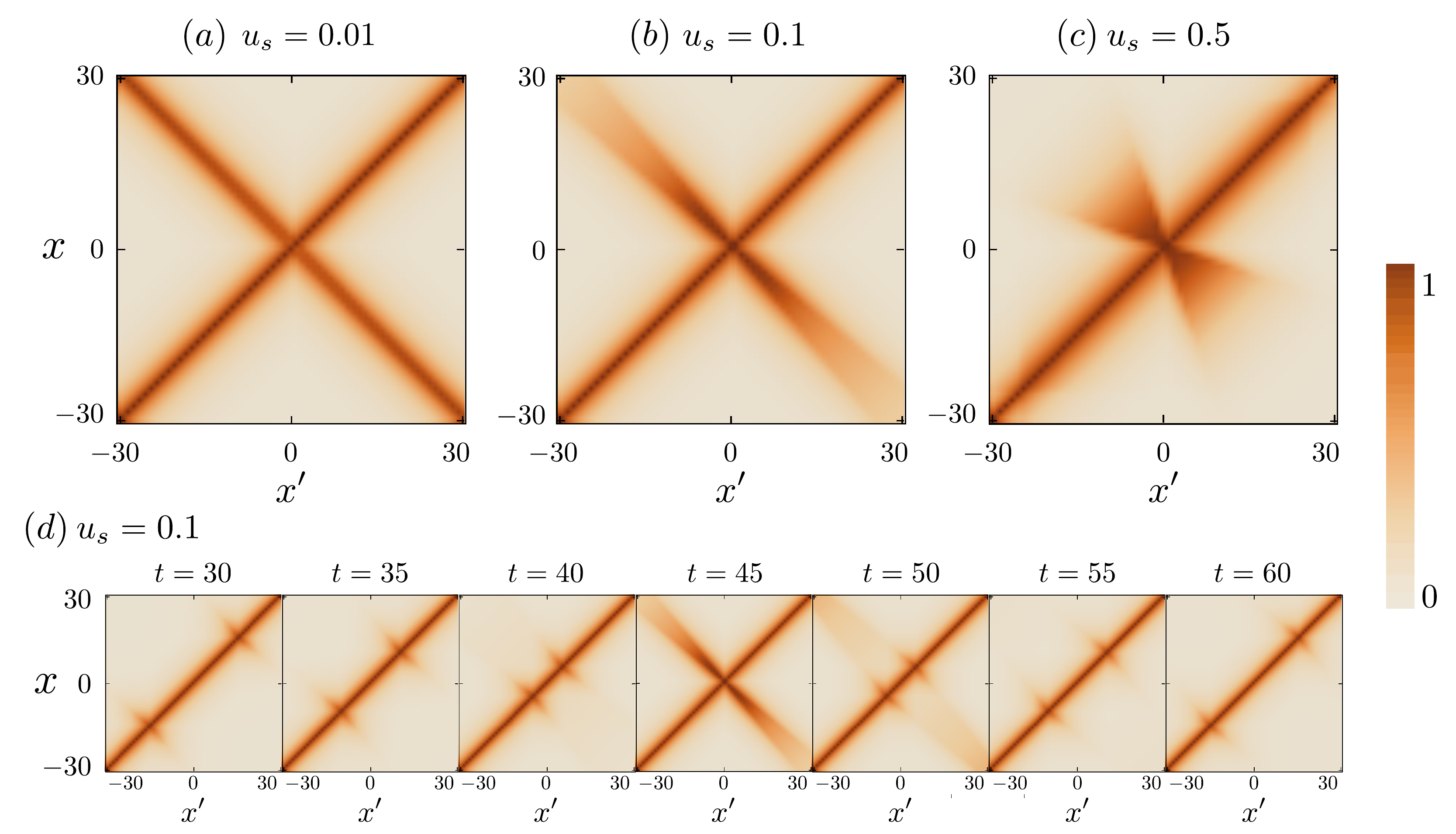}
\caption{Boundary Conditions used: $\partial_t \phi(L/2) = 0$. The correlations $\avg{e^{-i (\phi(x) - \phi(x')) }}$ are plotted in a matrix format for $x,x' \in [-30 \xi_c , 30 \xi_c]$, system size $L = 100 \xi_c$, Luttinger parameter $K = \rho \xi_c/2 = 10$ at times $t_c (u_s) = L(1-u_s)/(2c)$ for which cross correlations are strongest in (a), (b) and (c) for different splitting velocities $v_s = u_s^{-1}$ (in units of $c$). In (d), the development of the cross-correlations is shown in time for $u_s = 0.1c$.}
\label{fig:cross}
\end{figure*}
\end{center}
However, an inspection of the decay of equal-time correlations, as in Fig.~\ref{fig:localtemp} (a), (b) and (c), reveals that the correlations are not, in fact, described by such a prethermal state---this indicates the failure of the GGE since the occupation of eigenmodes in our system is not sufficient to correctly capture correlations. The reason for this deviation can be explained by noting that our system also carries \emph{slow} time-dependent correlations (as compared to the modes $\avg{a_{k_1} a_{k_2}}$ which oscillate at a frequency given by the sum of the energy of the modes $k_1$ and $k_2$) which evolve as $\sim e^{(k_1 - k_2)t}$ with momenta $k_1 \neq k_2$ that are related to one another by relativistic Doppler-shift factors. (These terms come with the prefactor $\beta_{k_1 k'} \beta^*_{k_2 k'}$.) These time-dependent terms involve correlations of two different momentum modes and consequently, cannot be captured by the usual GGE which takes into account only the occupation of different momentum modes.  

There are yet two ways in which our system does appear thermal. First, by construction, we expect that time-averaged correlations (averaged over the entire quantum revival cycle, or timespan $t = L/c$) should be described by thermal correlations at temperature $T_f$---the time-average simply eliminates all time-dependent correlations which lead to deviation from a prethermal state. Indeed, we see from Fig.~\ref{fig:localtemp} (d), that the log-averaged~\footnote{\label{note1} Log-averaging of an observable $C$ (over different times) is defined as $\avg{C}_{t} = \text{exp} \left( \avg{\text{log}C (t)} \right)$; here $\text{log} C(t)$ is averaged (by usual arithmetic averaging) over all times. We log-average the expectation values of $C \sim e^{i \phi(x,t) - i \phi(x',t)}$ at different times, because time-averaging the phase difference $\phi(x,t) - \phi(x',t)$ cancels time-dependent coherences, and not the exponential of this quantity, which is measured in experiments.}  value of $\avg{C(x_1,x_2)}_t = \avg{\avg{e^{i (\phi(x_1,t) - \phi(x_2,t))}}}_t$ over agrees with correlations of a thermal system at temperature $T_f$. However, such a notion of prethermalization has little value when considering a thermodynamically large system since it necessitates an extensively large number of measurements.  Second, and as we show below, we can yet explain the correlations in our system as being thermal, provided we restrict measurements to different regions of space-time. 

We note in passing that in a previous analysis~\cite{AgarwalChiral}, where system was considered to be infinitely large and undergoing a uni-directional split, the authors concluded that the system must enter a prethermal state with an effective temperature of $T_0 \gamma_s$ corresponding to a region that is influenced by both hot and cold waves. In our finite-system analysis, the waves coming from the center ($t = 0, x = 0$) of the splitter, and say, the right end ($t = - L/(2v_s)$, $x = L/2$) enclose a region of space-time wherein the infinite-size approximation is valid; within this region, the system does appear to be at a temperature $T_0 \gamma_s$ as predicted in the previous work. These regions (atop each half splitter) are enclosed within dashed lines in Fig.~\ref{fig:enerprop}. 

\subsection{Local prethermalization}

\begin{center}
\begin{figure*}
\includegraphics[width = \textwidth]{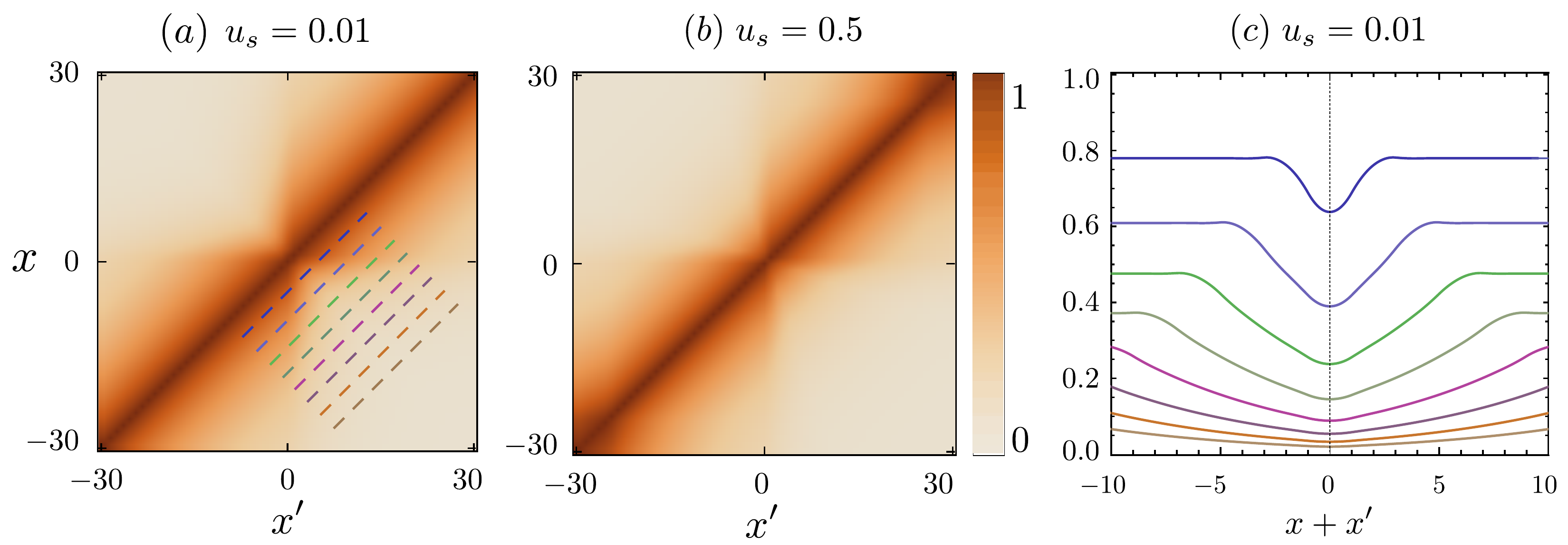}
\caption{Boundary Conditions used: $\partial_x \phi(L/2) = 0$. The correlations $\avg{e^{-i (\phi(x) - \phi(x')) }}$ are plotted in a matrix format for $x,x' \in [-30 \xi_c , 30 \xi_c]$, system size $L = 100 \xi_c$, Luttinger parameter $K = \rho \xi_c/2 = 40$ at times $t_c (u_s) = L(1-u_s)/(2c)$ for which cross \emph{anti}-correlations are strongest in (a) and (b) for two different splitting velocities $v_s = u_s^{-1}$ (in units of $c$). In (c), correlations (from top to bottom) along the cuts (dashed lines; away from diagonal) seen in (a) are plotted (in the same color), showing a marked suppression of correlations at $x' = -x$. }
\label{fig:cross2}
\end{figure*}
\end{center}
As mentioned above, the system cannot reach a true prethermal state because of time-dependent correlations associated with two different momentum modes. However, our intuitive picture of the dynamics of the systems makes it clear that there may be different regions of space-time that can be locally thought to exist at a particular effective temperature depending on whether these regions are inundated by cold, or hot, or both cold and hot waves. We except that regions only influenced by cold (hot) waves must exhibit an effective temperature of $T_0 / \eta_R$ ($T_0 \eta_R$). Regions which are inundated by both cold and hot waves should exist at an effective temperature $[T_0 / \eta_R + T_0 \eta_R ]/2 = T_0 \gamma_s$. These expectations are supported by numerical simulations; see Figs.~\ref{fig:localtemp}. Moreover, from Fig.~\ref{fig:localtemp} (c), we see that one can even use these local effective temperatures to correctly predict the decay of correlations across regions described by different temperatures.

\section{Anomalous, `cross' correlations}
\label{sec:crosscorrsec} 

We now discuss another feature of our analysis that does not appear in the discussion of an infinite system: the presence of strong correlations between $\phi(x,t)$ and $\phi(-x,t)$ that result due to quantum coherent revivals. This feature also appears in experiments, although the precise relation of these experiments to our splitting protocol is not clear (see also the concluding discussions); here we discuss the conditions under which such `cross-correlations' appear in our protocol. (The terminology is adopted by the suggestive appearance of the correlations as plotted in Fig.~\ref{fig:cross}.)

The emergence of these cross correlations are easiest to analytically discuss for the instantaneous quench. For this case, $\beta^\epsilon_{k k'} =  2\pi \epsilon \delta(k'-k) (\omega'_k - k) A_k B_k/g$ while $\alpha^\epsilon_{k k'} = 2 \pi \delta(k'- k) (\omega'_k + k) A_k B_k/g$, and we have, analytically,  

\begin{align}
& \avg{(\phi(x,t) - \phi(x',t))^2} \bigg|_{v_s = \infty} = \nonumber \\
& \sum_{k, \epsilon} (2 N_k +1) \abs{u^\epsilon_{k} (x) - u^\epsilon_k (x') }^2 \nonumber \\
&- \sum_{k, \epsilon} 2 N_k \frac{\omega'_k + k}{\omega'_k - k} \abs{u^\epsilon_k (x) - u^\epsilon_k (x') }^2 \cos (2 k t) .
\nonumber \\
\label{eq:inst}
\end{align}

The factor $(\omega'_k + k)/(\omega'_k - k) \approx 1$ for $k \ll m$. Thus, the anomalous term is approximately equal in amplitude to the `thermal' term besides an extra factor of $- \cos (2kt)$. This has important consequences for long-time dynamics. In particular, if the boundary condition $\partial_t \phi_\pm(L/2) = 0$ is satisfied, then at $t = L/(2c)$, $\cos(2 k t) = -1$ for \emph{all} symmetric modes while $\cos (2 k t) = 1$ for \emph{all} anti-symmetric modes. As a result, at $t = L/(2c)$, symmetric modes appear to be populated at twice the usual temperature $T_0$, while anti-symmetric modes appear to be in the ground state. This imbalance yields the strong positive correlations between the phase at point $x$ and $-x$ (see Fig.~\ref{fig:cross}) and is clearly, a purely transient quantum revival phenomenon. Alternatively, if the boundary condition $\partial_x \phi_\pm (L/2) = 0$ applies, then at $t = L/(2c)$, $\cos(2 k t)$ assumes the opposite values for (anti-) symmetric modes and we find that the system temporarily has highly populated anti-symmetric modes and zero-temperature symmetric modes. This results in negative correlations between the phase at points $x$ and $-x$, as illustrated in Fig.~\ref{fig:cross2}. 

The correlations over the region $x \in [ -10 , 10]  \xi_c$ and $x \in[ 0, 50] \xi_c$ and time $t \in [0, 100] \xi_c/c$ for the two different boundary conditions are available as movies.

For the general case of finite $u_s$, these cross correlations emerge at a time $t \sim L (1-u_s) / (2c)$. This is borne out by numerics, but can also be gleaned from examining the $L \rightarrow \infty$ results for $\alpha_{k,k'}$ and $\beta_{k,k'}$ in Eqs.~(\ref{eq:alphabetathermo}): we note that these crosses are a re-phasing phenomenon in which the anomalous terms $\propto \alpha_{k_1,k'} \beta_{k_2,k'}$ acquire a phase factor  $e^{-i \eta_R (k_1 + k_2) t}$ that is $-1$ for all anti-symmetric modes and $+1$ for all symmetric modes. From Eqs.~(\ref{eq:alphabetathermo}), we notice that one such (anomalous) term arises for $\eta_R^2 k_1 = k_2 = k'$; inserting this condition into the dynamical phase $e^{-i \eta_R (k_1 + k_2) t}$ and requiring it to be $\pm 1$ depending on whether $k_1,k_2$ are symmetric/anti-symmetric yields the (earliest) time for re-phasing, $t = L (1-u_s)/(2c)$. 

Finally, we contrast the cross-correlations observed in our analyses and the model used to explain the experimental findings in Ref.~\cite{Jorgcross}. In our case, these correlations arise as recurrent, transient quantum phenomena that are generated by temporary imbalances in the population of symmetric and anti-symmetric modes. In Ref.~\cite{Jorgcross}, it is suggested that the system reaches, at long times, a steady state wherein there exists a similar population imbalance, and which generates the cross-correlations. We note that such steady-state population imbalances are not found in our analysis, as evident from Eq.~(\ref{eq:popu}) and may arise in experiments due to the effect of parabolic confinement of the condensate. 

\subsection{Light-cone spread of correlations}

\begin{center}
\begin{figure}
\includegraphics[width = 2.6 in]{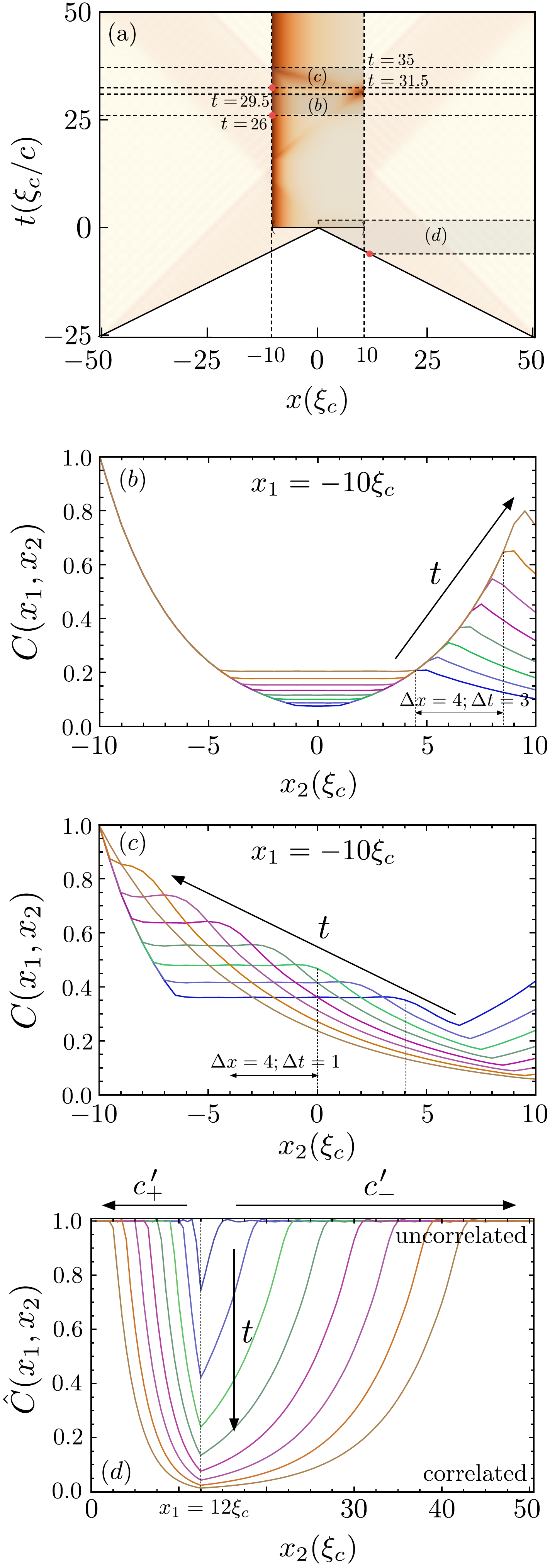}
\caption{(a) shows the correlation function $C(x_,  x_2)$, with $x_1 = -10 \xi_c$ (red dots) and $x_2 \in [-10,10] \xi_c$ over times $t \in [ 0 , 50] \xi_c/c$ and $u_s = 0.5c$. Non-thermal correlations are formed over supersonic `fronts' (see as dark orange lines in the region) that move at speeds $c'_\pm = 2 c / (1 \pm u_s/c)$. (b) and (c) examine the emergence and fading, respectively, of these non-thermal correlations in the region $x \in [ -10 \xi_c , 10 \xi_c]$ at times that are highlighted in (a), and verify the speeds $c'_+ = 4/3 c$ and $c'_- = 4 c$, as expected for $v_s = 2 c$, or $u_s = 0.5 c$. (d) provides another confirmation of the supersonic spread of correlations by examining the change of \emph{normalized} correlations $\hat{C} (x_1,x_2)$ (explained in the main text) over times $t \in [-6 \xi_c, 2 \xi_c]$, and $x_1 = 12 \xi_c$ (red dot), as shown in (a). Note $\hat{C} (x_1,x_2) = 1$ when points $x_1$ and $x_2$ are uncorrelated and  $< 1$ otherwise. System parameters are as in Fig.~\ref{fig:enerprop}. Boundary Conditions used: $\partial_t \phi(L/2) = 0$.}
\label{fig:lightcone}
\end{figure}
\end{center}

\begin{center}
\begin{figure}
\includegraphics[width = 2.6 in]{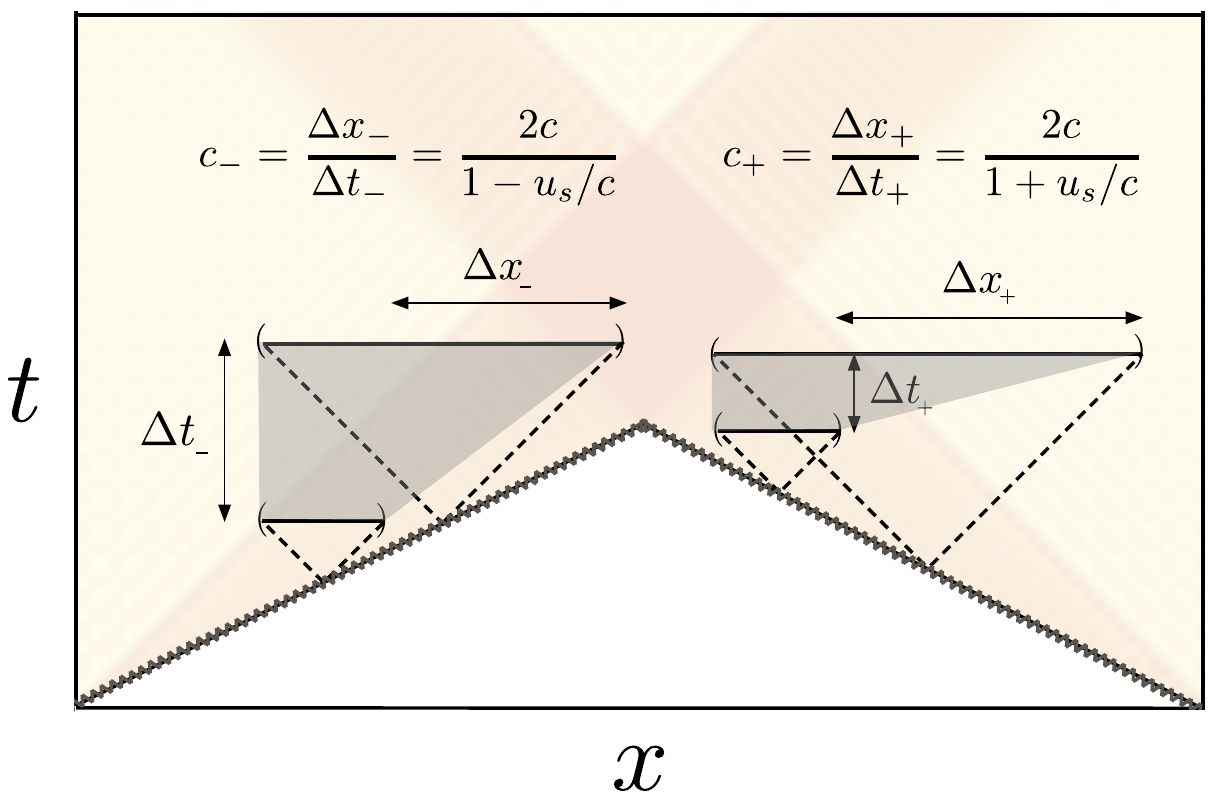}
\caption{The calculation of $c_-$ and $c_+$, the speeds that govern the spread of correlations in the system is illustrated. Pairs of oppositely traveling quasiparticles (dashed lines) are emitted by the splitter and entangle the system over the distance covered by them. The velocity of this spread is clearly $2 c$ if the splitting is instantaneous. The spread of correlations can happen faster or slower than $2c$ in our system since the splitting front is itself moving.}
\label{fig:lightconeexplain}
\end{figure}
\end{center}

We now provide an argument for the spread of such anomalous correlations due to revival and dephasing, and numerically justify our findings in Figs.~\ref{fig:lightcone} (b) and (c). We note that, at every point immediately after the split, the phase is close to zero, and can be therefore, thought to have long-range coherence. As we will show, the speed at which, thermal-looking correlations develop after this splitting process can give an indirect confirmation of the sound velocity $c$ and the speed of splitting $v_s$ in the system. This phenomenon has been examined experimentally for the instantaneous splitting case~\cite{Langen}; examining two-point correlations, it was found that $C (x_1, x_2, t)$ appears to take its exponentially decaying, or thermal form $e^{ - \text{const.} |x_1 - x_2|}$, for $x_2 <= x_1 + 2 c t$, where $t$ is the time measured after the splitting has occurred. Thus, the correlations `spread' at speed $c' = 2c$. This may be understood by noting that two quasi-particles originating from a common source (in space-time) and traveling in the opposite directions, disturb the initially perfectly correlated phased at points $x_1$ and $x_2$ in time $|x_1 - x_2|/ (2 c)$. 

The argument clearly needs to be revisited in our system given that the splitting, which generates the excited quasiparticles, occurs at different times at different points in space. This is illustrated in Fig.~\ref{fig:lightcone}, where we see the spread of correlations from the left splitter and right splitter spreading (to the right) at different speeds. The computation of these speeds is illustrated schematically in Fig.~\ref{fig:lightconeexplain} by showing how initial (and consequently, revival) correlations dephase (rephase) due to quasiparticles, that originate at the splitter, and connect points at distance $\Delta x_\pm$ in time $\Delta t_\pm$. These speeds can be evaluated straightforwardly to be $c'_\pm = 2c / (1 \pm u_s)$ and are both supersonic; $c'_+ < 2 c$, $c'_- > 2 c$. We note that such dephasing or revival is not bounded by light-cone physics which limits response functions, and can therefore occur at these supersonic speeds. For the case when the splitter moves at twice the speed of sound, $v_s = 2 c$ or $u_s = 0.5c$, as also in Figs.~\ref{fig:enerprop} and~\ref{fig:localtemp}, these speeds evaluate to $c'_+ = 4/3 c$ and $c'_- = 4c$. We verify in Figs.~\ref{fig:lightcone} (b) and (c) that these are indeed the speeds at which quantum revival and dephasing is observed. A more direct affirmation of these velocities that determine the spread of correlations is found by examining $\hat{C} (x_1, x_2) = \avg{e^{i \phi(x_1)}}\avg{e^{-i \phi(x_2)}} / \avg{e^{i \phi(x_1) -i \phi(x_2)} }$ which is $1$ if the phase $\phi(x_1, t)$ and $\phi(x_2, t)$ are uncorrelated and $< 1$ otherwise. Fig.~\ref{fig:lightcone} (d) examines the change of $\hat{C} (x_1,x_2)$ over time for fixed $x_1$ and verifies that initial correlations spread at speeds $c'_\pm$. 

Finally, we note that the speed at which the correlations change from one thermal form to another, occur at speed $c$, as is clear from Fig.~\ref{fig:enerprop} by examining the movement of the quantum heat waves. 

Movies illustrating graphically the change of correlations over time are available separately. 

\section{Conclusions}

We examined the problem of dephasing between two halves of a condensate that is split along a bi-directional supersonic trajectory. Following the discussion in Refs.~\cite{Bistritzer,Takuya,AgarwalChiral}, we modeled the phase difference as a Luttinger liquid and described the quantum quench as a transition from a Luttinger liquid with an additional mass to one without this mass. The mass effectively ties the relative phase fluctuations down to zero prior to the quench (bar quantum fluctuations) and eliminating it leads to the creation of left- and right-moving sound waves each carrying an energy of the order of the chemical potential. In the laboratory frame, these fronts appear to be Doppler-shifted. Equivalently, while the waves appear to be at an effective temperature $T_0$ in the Lorentz-boosted frame, they appear at a temperature $T_0/\eta_R$ and $T_0 \eta_R$ in the laboratory frame, depending on the direction of their motion relative to the moving quench boundary. Even though these waves cannot relax themselves, they lead the system into a state where time-averaged correlations can be described by a single effective temperature $T_0/ \gamma_s$, where $\gamma_s > 1$ is a Lorentz-dilation factor. Thus, when the splitting velocity matches the sound velocity, we find that the system is effectively at zero-temperature even after the global quench; all the energy in fact, gets dumped into two extremely hot sonic booms emanating from either (left and right) splitting front, that is infinitely narrow. 

We further found that, our intuitive picture of hot and cold waves provides a prescription to describe local correlations in regions of space-time as approximately thermal, with an effective temperature that depends on whether the space-time region is inundated by cold, hot or both cold and hot waves. Moreover, these local temperatures never agree with the temperature $T_0 / \gamma_s$ corresponding to the actual distribution of the Luttinger Liquid bosons. This shows that the non-equilibrium state we find in our system fails to be described by a Generalized Gibbs Ensemble, and correlations between bosons of different eigenmodes continue to remain relevant at all times. Thus, our system provides a remarkable example which does not comply with the GGE even at long times, and yet, the system appears to have thermal correlations; moreover, the precise temperature setting these correlations depends on the \emph{extensive} region of space-time over which measurements are conducted. 

Our analysis also uncovered certain `cross' correlations that appear ephemerally at late times. These correlations occur due to an imbalance in the population of symmetric and anti-symmetric modes and are superficially similar to those observed in experiments~\cite{Jorgcross}. In the present experiments~\cite{Gring,Jorgcross}, which study an instantaneous quench, it is likely that the ends of the condensate that are at a lower density, are split earlier, thus mimicking our quench protocol. However, these experiments also employ parabolic traps whose effects we do not consider and which may explain the \emph{study-state} imbalance in the population of long-wavelength symmetric and anti-symmetric modes observed~\cite{Jorgcross}. The parabolic shape will also introduce additional dephasing due to the presence of different local sound velocities and an extension of our analysis to such traps would be interesting. At present, we expect systems of ultra-cold atoms trapped in flat-band potentials~\cite{HadzibabicUniformBose,mukherjee2015fermi}, besides arrays of Josephson Junctions~\cite{HavilandArray,UstinovJJArray}, and ion traps~\cite{blatt2012quantumsimulation,MBLTrappedIonMonroe} to be viable candidates for experimentally probing the Luttinger-liquid physics we have considered in this work.  

Finally, we note that our picture of Doppler-shifted hot and cold waves is very general and should allow us to understand non-equilibrium dynamics following more general supersonic perturbations in finite systems exhibiting emergent Lorentz symmetry. We anticipate that such generalizations may lead to richer, more novel dynamical behavior, and further our understanding of the emergence of ergodicity in near-integrable and integrable quantum systems.

\section{Acknowledgements} 

We thank Anatoli Polkovnikov for useful discussions. The authors acknowledge support from Harvard-MIT CUA, NSF Grant No. DMR-1308435, AFOSR Quantum Simulation MURI, ARO MURI on Atomtronics, ARO MURI Qusim program, and AFOSR MURI Photonic Quantum Matter. KA acknowledges support from DOE-BES Grant No. DE-SC0002140. EGDT acknowledges support from the Israel Science Foundation, Grant No. 1542/14. JS acknowledges support from the ERC advanced grant Quantum Relax.


%


\end{document}